\documentclass[%
 aip,
 amsmath,amssymb,
reprint,%
]{revtex4-2}
\usepackage{xcolor} 
\usepackage{graphicx}
\usepackage{dcolumn}
\usepackage{bm}

\begin{document}

\preprint{AIP/123-QED}

\title{Dispersive measurement of a semiconductor double quantum dot via 3D integration of a high-impedance TiN resonator}

\author{Nathan Holman}
\affiliation{Department of Physics, University of Wisconsin-Madison, Madison, WI 53703, USA}
\email[Corresponding Author:~]{maeriksson@wisc.edu}

\author{D. Rosenberg}
\affiliation{MIT Lincoln Laboratory, 244 Wood Street, Lexington, MA 02421}

\author{D. Yost}
\affiliation{MIT Lincoln Laboratory, 244 Wood Street, Lexington, MA 02421}

\author{J.L. Yoder}
\affiliation{MIT Lincoln Laboratory, 244 Wood Street, Lexington, MA 02421}

\author{R. Das}
\affiliation{MIT Lincoln Laboratory, 244 Wood Street, Lexington, MA 02421}

\author{William D. Oliver}
\affiliation{MIT Lincoln Laboratory, 244 Wood Street, Lexington, MA 02421}
\affiliation{Research Laboratory of Electronics, Massachusetts Institute of Technology, 77 Massachusetts Avenue, Cambridge, MA 02139}


\author{R. McDermott}%
\affiliation{Department of Physics, University of Wisconsin-Madison, Madison, WI 53703, USA}

\author{M.A. Eriksson}
\affiliation{Department of Physics, University of Wisconsin-Madison, Madison, WI 53703, USA}

\date{\today}

\begin{abstract}
Spins in semiconductor quantum dots are a candidate for cryogenic quantum processors due to their exceptionally long coherence times. One major challenge to scaling quantum dot spin qubits is the dense wiring requirements, making it difficult to envision fabricating large arrays of nearest-neighbor-coupled qubits necessary for error correction. We describe a method to solve this problem by spacing the qubits out using high-impedance superconducting resonators with a 2D grid unit cell area of $0.16~\text{mm}^2$ using 3D integration. To prove the viability of this approach, we demonstrate 3D integration of a high-impedance TiN resonator coupled to a double quantum dot in a Si/SiGe heterostructure. Using the resonator as a dispersive gate sensor, we tune the device down to the single electron regime with an SNR = 5.36 limited by the resonator-dot capacitance. Characterization of the dot and resonator systems shows such integration can be done while maintaining low charge noise metrics for the quantum dots and with improved loaded quality factors for the superconducting resonator ($Q_L = 2.14 \times 10^4$), allowing for high-sensitivity charge detection and the potential for high fidelity 2-qubit gates. This work paves the way for 2D quantum dot qubit arrays with cavity mediated interactions.

\end{abstract}

\maketitle

\section{Introduction}
One major challenge for noisy intermediate-scale quantum (NISQ)-era superconductor and semiconductor based qubit systems lies in the wiring interconnect problem.\cite{Preskill2018,Franke2019} Unlike classical processors, where of order $10^3$ signal lines operate $10^9$ transistors, all solid state quantum bit platforms require at least one independent control line per qubit. In cryogenic solid state platforms, Josephson effect based qubits, such as the transmon, capacitively shunted flux, or fluxonium qubit, have a modest overhead of 1-3 lines per qubit for control and readout .\cite{Schreier2008,Barends2013,Yan2016,Nguyen2019} Semiconductor spin qubits in accumulation mode gate-defined quantum dots require between 9-13 control lines per qubit to form the necessary electrostatic environment, with anywhere from 1-5 of those lines requiring $\gtrsim 1$~GHz bandwidth to perform high fidelity qubit control operations. \cite{Yoneda2018,zajac2018, Reed2016, Andrews2019, Huang2019,Xue2019} In order to implement the surface code for quantum error correction, a 2D grid of $N\times N$ nearest neighbor coupled qubits is minimally necessary.\cite{Fowler2012} For semiconductor spin qubits, this requirement becomes an issue of great practical concern, as the need to fabricate dense networks of sub-100~nm sized gate electrodes to form a large array of coupled spin qubits will quickly become a major wire routing and interconnect engineering challenge for any all-semiconductor approach with current proposals requiring several technical innovations before becoming feasible.\cite{Veldhorst2017}

 Fortunately, circuit quantum electrodynamics (cQED) provides a framework in which the extreme wiring density requirements for quantum dot spin qubit processors may be alleviated by using compact high-impedance superconducting resonators to mediate long range 2-qubit interactions, providing room for the necessary wiring to form the individual qubit structures. Recent work has shown spin-cavity architectures can reach the strong coupling regime for charge, valley-orbit, and spin degrees of freedom as well as facilitate interactions between other spins or superconducting qubits.\cite{Mi2017,Mi2018_valley,Mi2018_spin,Samkharadze,Borjans2020Interaction,Scarlino2019,Landig2019} Using cavity mediated two qubit gates reduces the array problem to fabricating $N\times N$ copies of single qubit structures (linear dot arrays) which is routine in the lab. 
 
 \par Vertical integration further ameliorates the wiring density problem by allowing some large components to be placed vertically ``off-chip" in a 3D architecture. This approach uses two or more dies and thermomechanical bonding to create electrical contacts between the dies using indium bumps.\cite{Rosenberg2017,Foxen2017} Currently, 3D integrated circuits are the workhorse wiring solution for high qubit count superconducting quantum processors, facilitating the path to demonstrations of quantum supremacy and large quantum volumes with superconducting processors.\cite{Arute2019,Jurcevic2020} In this paper, we demonstrate the viability of the 3D integration approach that allows for a Si/SiGe double quantum dot (DQD) to be coupled to a vertically integrated high-Q, high-impedance TiN resonator used for dispersive gate charge sensing without degradation in performance of either component.
 

\section{Results}

\par
\begin{figure}[ht]
\includegraphics[width=\columnwidth]{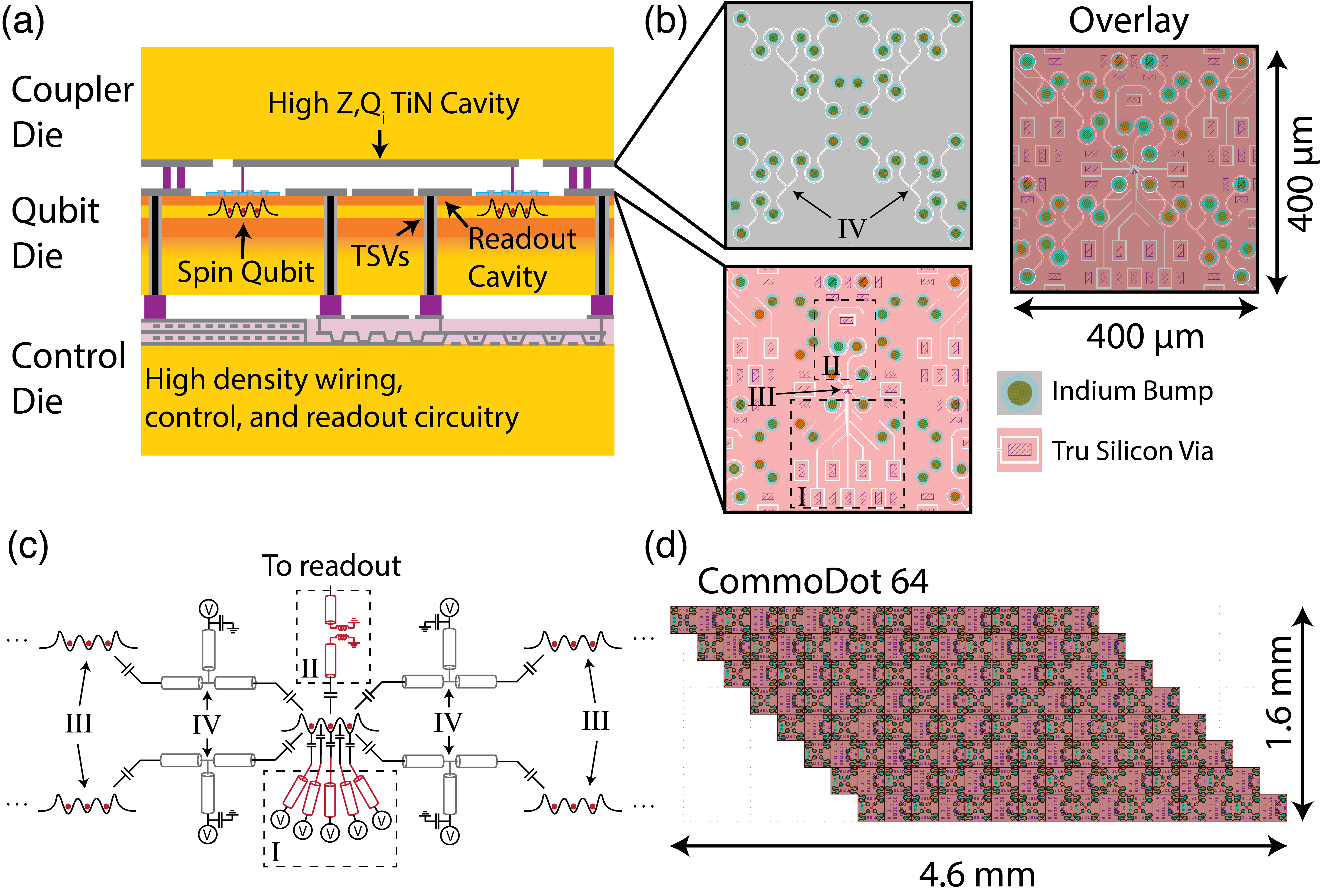}
\caption{\label{fig:Vision} (a) Diagram of a cQED spin qubit processor utilizing a 3-tier stack. (b) Layout of an exchange only qubit cQED processor unit cell with a $0.16~\text{mm}^2$ footprint. Top die consists of four $\lambda/2$ high-impedance resonators with $\lambda/4$ voltage taps. Base die consists of nine control lines (I), two readout resonators (II), and two qubits (III with one centered, $1/4$ in each corner). Control and readout lines are routed to the backplane of the die using through-silicon vias (TSVs) with a $10\times20~\mu\text{m}^2$ footprint.\cite{Yost2020} (c) A simplified circuit schematic with the elements in red on the qubit die and the elements in gray on the coupler die. (d) A 64 qubit processor tiled out from the base tile design fitting in a $4.6 \times 1.6~\text{mm}^2$ area.}
\end{figure}

 In Figure \ref{fig:Vision}(a) we illustrate a 3-tier stack integration scheme that is conceivable using state of the art 3D integration methods. \cite{Rosenberg2017,Rosenberg2019,Yost2020} The base die (the superconducting multichip module or SMCM) serves as a routing layer for high density control wiring, possibly in conjunction with cryogenic CMOS, superconducting amplifiers or photodetectors used for qubit biasing, manipulation, and readout. \cite{Bardin2019,Macklin2015,Opremcak2018} Superconducting through-silicon vias (TSVs) connect the qubit die to the complex control and readout layer using indium bump bonds to create electrical contact.\cite{Yost2020} Low-Q readout resonators are integrated onto the qubit chip as they are more tolerant to higher internal losses resulting from multilayer processing necessary to form the spin qubit structures.  The top die consists of a network of high-impedance, low loss resonators used for cavity mediated two-qubit interactions between distant quantum dot qubits. 

\par Figure \ref{fig:Vision} (b) shows the CAD design of a lattice unit cell consisting of two qubits, four coupling bus resonators (IV, grey die), and two readout resonators (II, red die) fitting in a $0.16~\text{mm}^2$ top down area across two die. Nine TSVs (hatched purple, in Fig. \ref{fig:Vision} b) route bias lines for each triple quantum dot vertically into each unit cell. Figure \ref{fig:Vision} (c) provides a simplified circuit schematic of the unit cell with the elements on the qubit die colored in red and the elements on the coupler die colored in grey. Using this design Figure \ref{fig:Vision} (d) shows how a 64 qubit processor can be tiled out in a $4.6 \times 1.6~\text{mm}^2$ footprint. This design builds upon demonstrated technology and assumes nominal materials parameters that are possible for high kinetic inductance films (e.g. quantum dot gate design, TSV or indium bump lithographic dimensions, sheet kinetic inductance, etc.). \cite{zajac2016,Yost2020,Rosenberg2017,Niepce2019,Shearrow2020} We emphasize here this design requires fabrication process development to ensure thermal budget compatibility between the TSV process and the quantum dot fabrication.

\par
\begin{figure*}[ht]
\includegraphics[width=2\columnwidth]{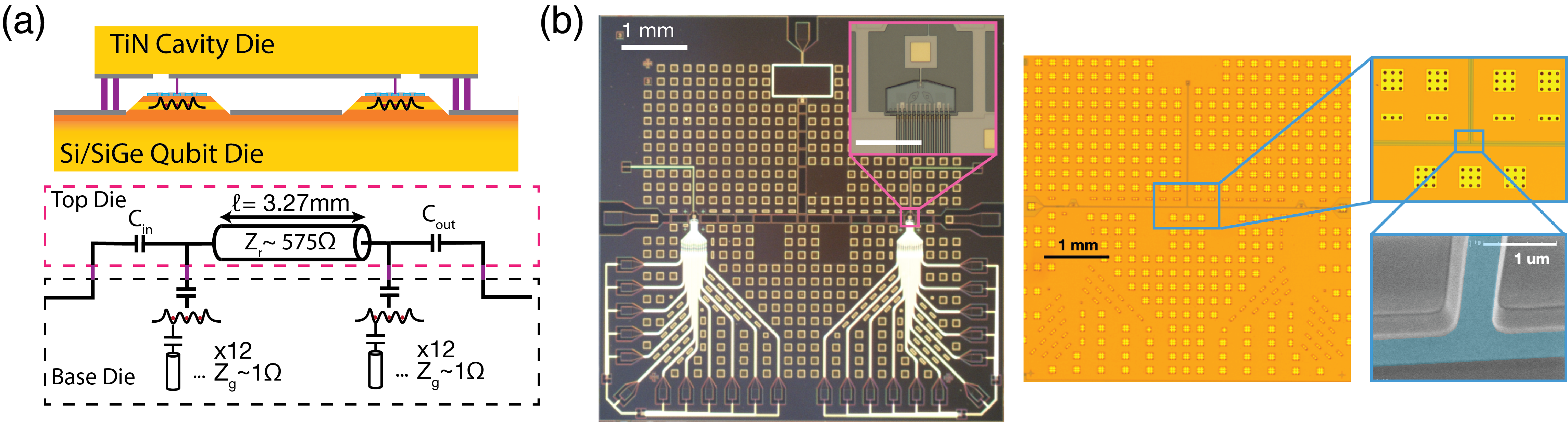}
\caption{\label{fig:Sample} (a) Upper panel shows a cross section of the 2-tier stack consisting of the high-impedance resonators and Si/SiGe qubit die. Lower shows a simplified circuit diagram for the device. (b) Left panel shows a dark field optical microscope image of a completed qubit die chip. Gold squares are underbump metalization consisting of Ti/Pd/Ti/Pt/Au used to facilitate galvanic contact between the qubit and coupling resonator die. Inset: Si/SiGe mesa and resonator landing pad prior to e-beam lithography of the quantum dot gates (scale = $100~\mu \text{m}$). Right panel shows optical and scanning electron micrographs of the TiN resonator die.}
\end{figure*}

 For our experiment, we limit the architecture to the 2-tier stack consisting of a quantum dot qubit die and high-impedance cavity coupler die as illustrated in Figure \ref{fig:Sample} (a). Optical and scanning electron micrograph images of the two dies prior to bonding can be seen in Figure \ref{fig:Sample} (b). The base die consists of a Si/SiGe heterostructure in which accumulation-mode, gate-defined quantum dots are formed. The cQED coupler die consists of a high kinetic inductance, high-impedance TiN $\lambda/2$ coplanar waveguide resonator. To minimize parasitic photon loss from coupling of the resonator to the 24 dot bias leads, we fabricate low impedance control lines using buried coplanar waveguides (BCPWs) with a characteristic impedance of $Z_{g}\approx1~\Omega$. The low characteristic impedance ameliorates unwanted loss out the dot leads.\cite{Holman2020} 

\par The BCPWs consist of a coplanar waveguide (CPW) made from niobium, a $\text{SiO}_2$ dielectric layer deposited over the center conductor, and a capping niobium layer connecting the two ground electrodes. These BCPWs can be seen in the lower half of the Si/SiGe die in Figure \ref{fig:Sample} (b). The addition of the capping ground plane over the control wiring suppresses cross capacitances between control lines as well as the parasitic leakage capacitance to the resonator. The cavity coupler die consists of a $\lambda/2$ TiN CPW with a $\lambda/4$ segment shunted by a large parallel plate capacitor on the quantum dot die used for low frequency voltage biasing.\cite{Holman2020} TiN was chosen for its high kinetic inductance ($L_k$) and high internal \emph{Q} allowing for $Z_r > 1~\text{k}\Omega$ while maintaining $Q_i > 10^5$. \cite{Coumou2013,Shearrow2020}

To push into the high-impedance regime, we use a 50~nm thick TiN film with a 500~nm wide center pin, and 10.25~$\mu$m gap, as shown in Figure \ref{fig:Sample} (b). Using COMOSL Multiphysics to simulate the 3D structure's influence on the CPW line capacitance and the nominal kinetic inductance of the TiN film ($L_k \approx 9~\text{pH}/\square$) we estimate a characteristic impedance of $Z_r = 575~\Omega$ with target fundamental frequency of 4~GHz. We find these estimates are in good agreement with the experimentally measured range of frequencies 3.985-4.115~GHz from several different samples with nominally identical resonator dimensions. Galvanic contact between both dies is facilitated by underbump metal pads made from Ti/Pt/Au on the resonator die and Ti/Pd/Ti/Pt/Au on the quantum dot die which do not form native oxides. To avoid unwanted damping caused by power dissipation of the microwave currents in the resonator to the normal metal pads, they are placed at voltage antinodes and at the end of the quarter wave DC tap where nominally no microwave currents exist for the $\lambda/2$ mode, thus preserving the high internal quality factor.\cite{Beck2016,Rosenberg2017}

\par
\begin{figure}[ht]
\includegraphics[width=\columnwidth]{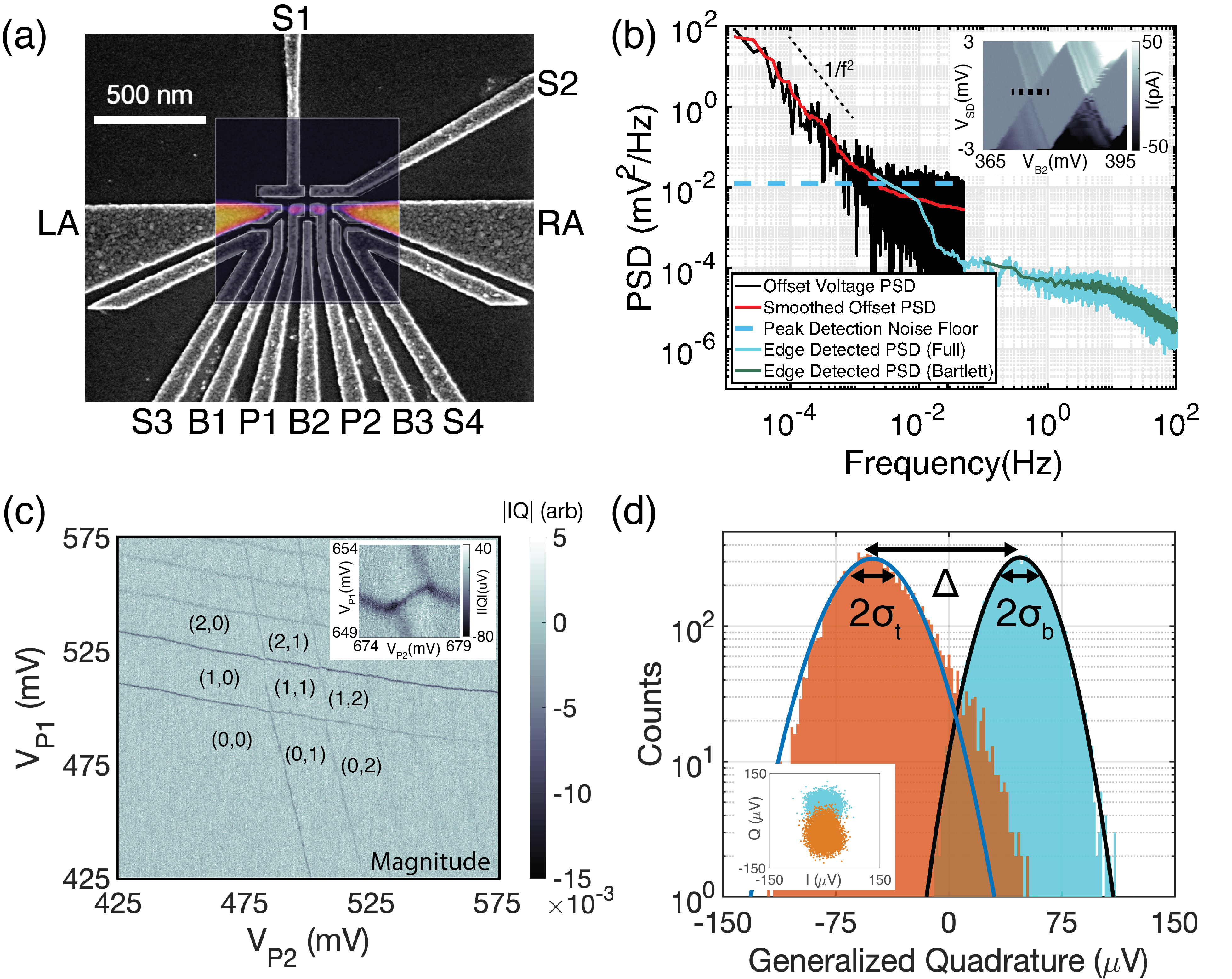}
\caption{\label{fig:NoiseAndReadout} (a) An SEM of the single layer double quantum dot (DQD) aluminum gate structure with the resonator connected to S1. Overlaid is a Thomas-Fermi simulation of a DQD in the few electron regime. (b) DC transport, gate referred noise measurements in the device. The black trace is acquired using a peak location monitoring method over a 21 hour period (red is smoothed using a moving mean). The teal and green traces are extracted by measuring current fluctuations at the maximum first derivative point of a Coulomb blockade peak for 510~s. Inset: Bias spectroscopy of a single electron quantum dot intentionally formed under B2 used to obtain the noise data. (c) Dispersive gate charge sensing of a double quantum dot formed under P1 and P2 down to single electron occupancy using the high-impedance TiN resonator as a dispersive gate readout. Inset: A tunnel coupled many electron double quantum dot with $n_{L,R} \approx 10$. (d) Generalized IQ quadrature histograms showing the visibility of a charge transition with a drive power of approximately $-95$~dBm on chip with an $\text{SNR}\approx 5.36$. Solid lines are Gaussian fits to the data. Inset: Demodulated IQ blobs used to generate the histograms.}
\end{figure}

 For these data a single layer aluminum gate stack for forming a double quantum dot is used. \cite{Samkharadze,Zheng2019} Underneath the gate stack is an ultra thin ($\sim1.7$~nm thick) $\text{SiO}_2$ layer grown by nitric acid oxidation of silicon (NAOS). \cite{Koybashi2010} A scanning electron micrograph with an overlaid Thomas-Fermi simulation of the induced electron gas can be seen in Figure \ref{fig:NoiseAndReadout}(a). The two dimensional electron gas (2DEG) is induced in an 8.6~nm thick silicon quantum well made from 800~ppm $^{28}\text{Si}$ approximately 42~nm below the surface. As designed, the gates labeled (P1,P2) are intended for accumulating a double quantum dot and the gates labeled (B1:B2:B3) serve to tune the various tunnel barriers. The other gates labeled (S1:S2:S3:S4) help corral the charges to upper portion of the plunger gates and mitigate unwanted transport currents. The superconducting resonator has a galvanic connection to the gate S1 as a single layer variant of a split-gate coupler design which serves to decouple the cavity pin voltage from the neighboring quantum dot easing tuning constraints for two qubit samples. \cite{Borjans2020} Tuning the device as intended, a double quantum dot (DQD) can be formed and sensed under P1 and P2 using the TiN resonator as shown in Figure \ref{fig:NoiseAndReadout}~(c). In the many electron regime the DQD has tunnel rates comparable to the resonator frequency resulting in a visible interdot transition shown in the inset of Figure \ref{fig:NoiseAndReadout}~(c). 

\par A third dot can be formed under B2 by repurposing P1 and P2 as barrier gates. Bias spectroscopy of the B2 dot's last electron Coulomb diamond is shown in the inset of Figure \ref{fig:NoiseAndReadout}~(b). For each dot, we use bias spectroscopy to extract the lever arm of the nominal chemical potential gate and find a typical lever arms ranging between $\alpha = 0.20-0.25$ eV/V. In general, we observe good electrical confinement for each dot with charging energies ranging from $E_c = 3-5$~meV (3~meV for the B2 dot and 5~meV for the P1/P2 dots) with orbital energies $E_{orb} \approx 1$~meV in the few to single electron regime. Additional parasitic quantum dots can be observed under certain bias conditions, likely due to uncontrolled accumulation of 2DEG under the gate electrodes\cite{Xanthe2019} and are a source of low frequency instabilities in the device. These instabilities prevented tunnel coupling at frequencies comparable to the resonator frequency in the single electron regime.

\par To measure the level of charge noise the quantum dots experience at low frequencies we use two methods: Coulomb blockade peak location monitoring and voltage-to-current transduction. \cite{Freeman2016} Using both methods allows for noise to be characterized over 7 orders of magnitude from $10^{-5} - 10^{2}$~Hz. The first method is performed by repeatedly sweeping the plunger gate voltage (along the dashed line in the inset of Fig. \ref{fig:NoiseAndReadout}(b)) of the dot over the course of a day to monitor its location with a repetition rate of approximately 0.1~Hz. The data are then fit to a thermally broadened conductance peak defined by\cite{Beenakker}
\begin{equation}
I(V_g) = I_{\text{offset}} + \frac{A_0}{4 k_b T_e} \cosh{\left(\frac{\alpha (V_g-V_{\text{offset}})}{2 k_b T_e}\right)}^{-2}.
\end{equation}
From the fits the we extract a minimum electron temperature of $T_e \approx 200$~mK, peak current ($\frac{A_0}{4 k_b T_e}$), where $A_0$ is a fitting factor that depends on the biasing of the dot ($A_0 = V_{\text{SD}} 4 k_b T_e G_{\text{max}} = e^2 V_{\text{SD}}(\Gamma_r || \Gamma_l)$), and the peak offset voltage ($V_{\text{offset}}$). Using the extracted offset voltage for each scan, we generate a corresponding time series from which the power spectral density of the offset voltage can be computed as shown by the black (raw) and red (smoothed) traces in Figure \ref{fig:NoiseAndReadout}(d). We observe a $1/f^2$ power law dependence in the offset voltage at frequencies below 1~mHz similar to that observed in stadium style devices\cite{Struck2020}, characteristic of Brownian motion in the fluctuating variable.\cite{Krapf2018} The noise floor the measurement is limited by the ability to resolve voltage shifts smaller than the width of the peak with an approximate bound $\Delta V_g^2/\Delta t \approx 10^{-2}~\text{mV}^2/\text{Hz}$ where $\Delta V_g$ is the full width half max of the blockade peak and $\Delta t$ is the time between two successive 1D scans. Narrowing of the blockade peak by reduction in electron temperature provides a straightforward route to extending this method to higher frequencies without additional averaging, which is time consuming due to the very low frequency of the measurement.

\par To resolve offset voltage noise smaller than the size of the conductance peak, we use a second method where we measure fluctuations in the transport current at the point of maximum voltage to current transduction and compute the associated current noise power spectrum. \cite{Freeman2016,Connors2019} Assuming a linear transfer function between voltage and current we can convert the current noise power spectrum to a gate referred voltage power spectrum by computing 
\begin{equation}
S_V(f) = \bigg|\frac{dI}{dV_g}\bigg|^{-2}S_I(f),
\end{equation}
where $S_{V(I)}$ are the power spectral densities of the voltage (current) and $\frac{dI}{dV_g}$ is the derivative at the point along the blockade peak where the gate voltage was set for the data acquisition. Using this method we are able to extract the offset voltage noise spectrum between $2~\text{x}~10^{-3}$ - 100~Hz as shown by the teal trace in Figure \ref{fig:NoiseAndReadout} (b). A moving average filter is used to reduce statistical fluctuations in the data without distorting the overall shape of the signal.\cite{Yan2012}

\par The same data can be subdivided into 10~s traces where each trace's power spectrum can be averaged producing the averaged dark green spectrum from 0.1 - 100~Hz in Figure \ref{fig:NoiseAndReadout}~(b), in agreement with the smoothed data from the single time series. In most spectra from this device we observe a Lorentzian spectrum at low frequencies with a characteristic switching time between 1-10~Hz. The spectra measured have an amplitude at $f=1~\text{Hz}$ of $S_V(f) = 10 - 36~\mu\text{V}^2/\text{Hz}$ comparable with other work. \cite{Jones2019} Alternatively, one can cast in terms of chemical potential fluctuations: $S_{\mu}(f) = \alpha^2 S_V(f) = 0.6 - 1.5~\mu e\text{V}^2/\text{Hz}$, or in terms of offset charge\cite{Basset2014}: $S_c(f) = (e/E_c)^2 S_\mu(f) = 2.4 - 6\times10^{-2}~\text{m}e^2/\text{Hz}$
at $f=1~\text{Hz}$ which corroborates recent work that reducing the volume of deposited dielectrics above the quantum dots mitigates the effects of charge noise. \cite{Connors2019} The variation is somewhat dependent on the tuning of the device, with the lowest numbers corresponding to small source-drain bias $V_{\text{SD}}\leq50~\mu\text{V}$ (while maintaining $\frac{dI}{dV_g}>100~\text{pA}/\text{mV}$) and the relative amplitude of the low frequency switcher in this device.  The two methods measured noise amplitudes and exponents that are similar where they cross over around 1-10~mHz, demonstrating they are complementary methods to measuring noise in the quantum dot. Elimination of unwanted nearby 2DEGs through use of screening gates could eliminate the frequency instability and minimize parasitic switchers coupled to the intended dots. \cite{zajac2015}

To evaluate the resonator as a dispersive gate readout, we measure the the signal to noise ratio (SNR) for observing a tunneling event. We first tune a dot-lead transition to maximize the phase shift in the microwave tone at a fixed power of -95 dBm on chip. Next, we take $10^4$ measurements of the demodulated IQ voltages when the electron is biased in Coulomb blockade or tunneling with a 50~ms integration time sampled at $100~\text{kSa}/s$. We then rotate the IQ blobs to align along the Q quadrature and subtract the mean offset, as shown in the inset to Figure \ref{fig:NoiseAndReadout} (d). This maximizes the difference in the signal compared to integrating along the unrotated I or Q directions. As shown in Figure \ref{fig:NoiseAndReadout} (d), by taking a histogram the data along the Q quadrature we find the blockade peak voltage is well-described by a Gaussian process (solid line is a fit), while the tunneling peak undergoes a non-Gaussian process evidenced by the asymmetry of the peak and substantial deviation from the fit. This is possibly due to the low frequency switcher observed in the transport noise data causing the location of the peak location to telegraph voltage space. Curiously, the non-Gaussian shoulder is not present at lower drive powers, suggesting the process is stimulated from the microwave energy in the resonator (see supplemental Figure \ref{fig:SNR}). 
\begin{figure*}[ht]
\includegraphics [width=2\columnwidth]{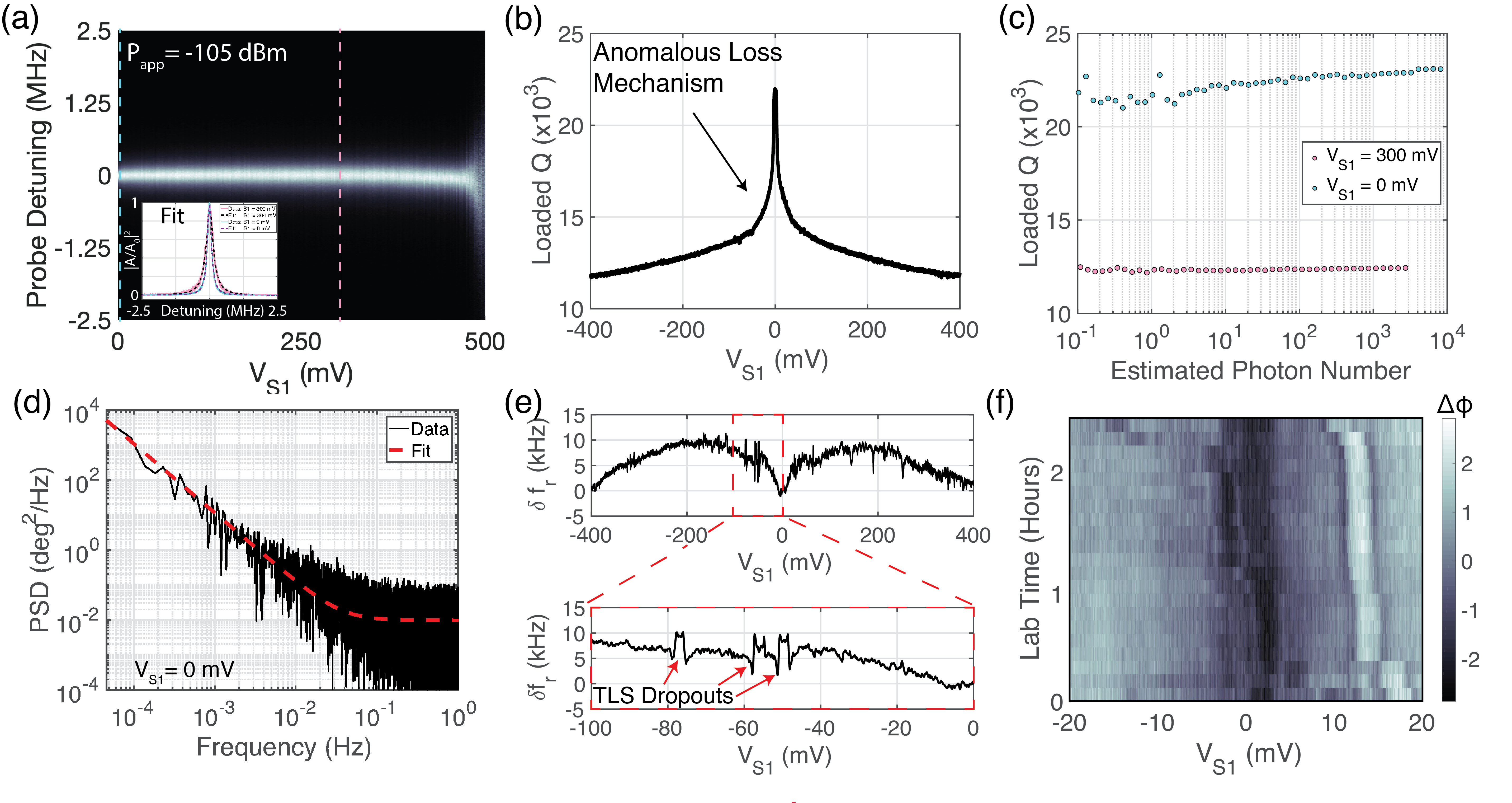}
\caption{\label{fig:NoiseQAndTLS} (a) Measurement of the resonator fundamental transmission spectra as a function of gate voltage. Inset: extracted Lorentzian fits to the teal and pink traces at $V_{S1} = 0$~mV and $V_{S1}=300$~mV respectively. (b) Extracted loaded quality factor as a function of gate voltage showing an anomalous decrease in the loaded quality factor upon voltage biasing the center pin. A narrow plateau in $Q_L$ occurs between $V_{S1} = \pm 2$~mV where the additional loss is less than the coupling quality factor.  (c) Power dependence of the loaded quality factor at 0~mV and 300~mV center pin bias. We extract a single photon internal quality factor of $Q_i\approx 3\times10^5$ consistent with radiative losses from the large CPW geometry.\cite{Sage2011} We estimate degradation of the internal quality factor to $Q_i\approx 3\times10^4$ at $V_{S1} = 300$~mV. (d) Phase noise power spectral density of the resonator with $V_{S1} = 0$~mV by probing the cavity on resonance with the dots empty. We extract a $1/f^\beta$ with $\beta = 1.98$ noise spectrum in the phase noise for frequencies below 0.1~Hz. (e) Upper: Resonator frequency shift ($\delta f_r = f_r(V_{S1} = 0) - f_r(V_{S1})$) as a function of voltage bias on the cavity pin. We observe a non-monotonic modulation in the center frequency. Lower: Zoom-in of a bias region in which multiple TLS-cavity interactions are observed. (f) Spectral motion of two TLS interacting with the TiN resonator over several hours.}
\end{figure*}

\par Defining SNR in terms of the separation of the demodulated peaks ($\Delta$) and the blockade peak standard deviation ($\sigma_b$), we find $\text{SNR} = \Delta/\sigma_b \approx 5.36$ which compares favorably to PCB integrated reflectometry methods while using somewhat lower powers.\cite{West2019} We note using $\sigma_b$ instead of $\sigma_t$ inflates the SNR by roughly $20\%$, as we empirically find the tunneling peak standard deviation ($\sigma_t$) is larger by approximately that amount. Compared to other resonator measurement techniques, which define SNR as the power SNR ($\text{SNR}_{P} = (\Delta/\sigma_b)^2$) we are notably much lower with $\text{SNR}_{P} = 28.7$ compared to $\text{SNR}_{P} > 1000$ for similar integration times.\cite{Stehlik2015,Zheng2019} The likely reason for the large difference is due to the substantially weaker coupling strength between the resonator and the quantum dots in our system. Based on S1 vs. P1(2) measurements, the lever arm of the resonator gate to the quantum dots is quite small ($\approx 0.05~\alpha_{P1(2)}$), causing the corresponding difference in the IQ signal during tunneling events to be substantially smaller than plunger coupled devices. Attempts to tune the coupling strength by tuning the resonator gate voltage\cite{Borjans2020} are hampered by a reductions in the loaded \emph{Q}, as discussed later in the text. We emphasize this issue is not intrinsic to the 3D architecture but rather a bug of the single layer gate layout resulting in poor placement of the quantum dots relative to the cavity electrode. Use of an overlapping gate architecture which has more precise placement of the quantum dots will substantially improve this aspect of the device performance. Additional optimizations such as using heterodyne detection with fast sampling DACs or quantum limited superconducting amplifiers can also improve the SNR through noise mitigation.\cite{Zheng2019,Macklin2015}


To characterize the effectiveness of our improved leakage suppression technique, we performed a systematic study of the quality factor with the device tuned in the (0,0) charge configuration while keeping the 2DEG reservoirs accumulated. Figure \ref{fig:NoiseQAndTLS} (a) shows the normalized $|S_{21}|^2$ response function of the TiN resonator as a function of the voltage bias on the resonator pin. At high bias ($V_{S1}>450~\text{mV}$), a substantial degradation in $Q_L$ is observed, due to induced 2DEG under the resonator gate substantially damping the cavity mode. Curiously, line cut comparisons of zero voltage bias (teal line) and sub-accumulation bias (pink line) show the linewidth of the resonator is substantially at finite voltage bias (see inset to Figure \ref{fig:NoiseQAndTLS} (a)). Performing a similar scan over a range $V_{S1} = \pm 400~\text{mV}$, we observe a dramatic change in the loaded quality factor with the magnitude of the applied electric field ($\propto|V_{S1}|$), as shown in Figure \ref{fig:NoiseQAndTLS} (b). The origin of the anomalous loss mechanism upon voltage biasing the resonator is unclear but likely cannot be attributed to the induced 2DEGs in under the accumulation gates LA or RA, as degradation occurs regardless of the sign of the bias. To extract the internal \emph{Q} ($Q_i$), we perform power sweeps at two voltage biases, shown in Figure \ref{fig:NoiseQAndTLS} (c). Assuming the high power limit at zero voltage bias is defined by the explicit coupling capacitances for probe and readout, we extract $Q_c = 2.31\times10^4$ with $Q_i (V_{S1} = 0~\text{mV}) \approx 3\times10^5$ and $Q_i (V_{S1} = 300~\text{mV}) \approx 3\times10^4$. To our knowledge, the extracted $Q_i$ at zero voltage bias is the highest measured $Q_i$ in a superconducting-semiconductor hybrid system.

It has recently been proposed that high-impedance resonators may exhibit lowered quality factors due to enhanced phase noise rather than true energy loss, due to fluctuations in the kinetic inductance from charge noise resulting in a corresponding frequency modulation.\cite{Sueur2018} To see if this was present, we measured the phase noise at low frequencies by probing the transmission phase on resonance (at time $t=0$) over the course of several hours. We compute the corresponding phase noise power spectral density, as shown in Figure \ref{fig:NoiseQAndTLS} (d). We observe a $1/f^2$ dependence of the phase noise PSD below 0.1 Hz, atypical of high \emph{Q} superconducting resonators.\cite{Barends2007,Gao2007} Additionally, when performing the voltage bias studies, we observed the cavity resonance frequency is a nonmonotonic function of the applied voltage bias to the center pin with its frequency modulation varying a total range of $14~\text{kHz}$, as shown in Figure \ref{fig:NoiseQAndTLS} (e). Several reproducible TLS-cavity crossings, inferred by shifts of the resonator frequency, are observed over the $\pm400$~mV tuning range explored. This suggests the effect is present, but due to their small magnitude ($\approx 3-5~\text{kHz}$) are insufficient to explain the factor of $\sim10$ reduction in $Q_i$ with application of voltage bias. We note these TLS are not fixed in location in voltage space and undergo time dependent spectral diffusion over hours-long timescales, illustrated by the data in Figure \ref{fig:NoiseQAndTLS} (f). At this time it is unclear if these defects originate from the quantum dot die or the TiN resonator die or if any of the noise between the two systems is correlated.

In summary, we have described a 3D integration approach to hybrid superconductor-semiconductor quantum dot processors, demonstrated dispersive charge sensing of a DQD in a Si/SiGe heterostructure in a 2-tier stack using a high $Q_L$ superconducting high-impedance TiN resonator. By careful engineering of the bias wiring impedance, loaded quality factors as high as $2.14\times10^4$ were demonstrated with estimated $Q_i \approx 3\times10^5$ with the quantum dot portion of the device accumulated. Despite low cavity-dot coupling, we still are able to charge sense the quantum dots with an $\text{SNR} = 5.36$, sufficient for device tune up in the single electron regime. Transport measurements of charge noise over seven orders of magnitude shows a cross-over to $1/f^2$ below 10~mHz. A similar power law spectrum is observed in the resonator phase noise along with small frequency tuning and signatures of TLS-cavity coupling upon voltage biasing the resonator gate. Establishing the origins and possible connection between transport measured charge noise and cavity phase noise in high-impedance resonators is a point of future work and may provide a useful tool for rapid characterization of gate dielectrics in quantum dot devices.

\section{Methods}
Devices are packaged via aluminum wire bonding in a hybrid PCB and metal enclosure designed to limit parasitic chip modes out to 20~GHz for non-2-tier stack samples. The devices are then loaded into a Leiden CF-450 dilution refrigerator with a base temperature of 50~mK. Substantial care is taken to ensure the sample, fridge wiring, and experimentalist remain grounded during the load to minimize electrostatic discharge (ESD) risk to the sample. Use of a grounded dissipative floor under the cryostat, grounding wrist straps, antistatic coats, tools, gloves, humidity control ($35-55$~\%RH), and ionizing fans do not appear to substantially mitigate the ESD risk to the sample in our experimental setup (load yield around 10~\%). Compared to standard $50~\Omega$ resonator devices we observe increased catastrophic failure (discerned by gate to gate shorts) from ESD likely due to the high resistance nature of the TiN resonator pin at room temperature resulting in damage during packaging or loading. Delamination of a completed 2-tier stack device shows no apparent ESD from the thermomechanical bonding process itself to the quantum dot gate stack, corroborating ESD is a packaging or loading issue rather than a fabrication problem. Passive on-chip ESD protection measures such as sub-degenerate phosphorous doping between bond pads or freeze out TiN resistors to chip ground \cite{Shearrow2020} would serve as a future routes to improving sample yield.

A detailed measurement setup and fridge wiring schematic is provided in the supplement. Transport current measurements are done using a battery-powered DL1211 transimpedance amplifier with $10^9$ gain and an effective bandwidth of $\approx 2$~kHz. The output signal is sent to a SR560 voltage preamplifier with unity gain and a 10~kHz cutoff two pole low pass filter and sampled by an NI-DAQ 6216 with sampling rates between $1-10$~kSa/s. Microwave characterization measurements of the resonator are done using an Agilent N5230A vector network analyzer. Homodyne detection of the double quantum dot via the resonator are done using an Agilent E8257D PSG Analog Signal Generator with a power level corresponding to approximately $-95$~dBm on chip ($\approx 1.7\times10^4$ intracavity photons). The signal is amplified by cryogenic and room temperature amplifiers, filtered, and then demodulated by a Marki 0416 IQ mixer. The demodulated DC voltages are filtered and sent to a pair of SR560 voltage preamplifiers at unity gain with a 100~kHz two pole low pass filter and are then sampled by an NI-DAQ 6216 at 100~kSa/s with 5~kSa per point for the data shown resulting in an $\text{SNR}\approx 5.36$.

\section{Acknowledgements}
We thank L.F. Edge for providing the Si/SiGe heterostructure used in this work. We acknowledge discussions on modeling 3D architecture structures in COMSOL with E. Leonard, M. Beck, and C. Liu. We acknowledge helpful discussions with J. Kerckhoff. Research was sponsored in part by the Army Research Office (ARO) under Grant Numbers W911NF-17-1-0274. We acknowledge the use of facilities supported by NSF through the UW-Madison MRSEC (DMR-1720415) and the MRI program (DMR–1625348). Work done at MIT Lincoln Laboratory was funded in part by the Assistant Secretary of Defense for Research \& Engineering under Air Force Contract No. FA8721-05-C-0002. The views, conclusions, and recommendations contained in this document are those of the authors and are not necessarily endorsed nor should they be interpreted as representing the official policies, either expressed or implied, of the Army Research Office (ARO) or the U.S. Government. The U.S. Government is authorized to reproduce and distribute reprints for Government purposes notwithstanding any copyright notation herein.

\section{Supplement}

\subsection{Device Fabrication Process}
The quantum dot base dies are fabricated using a 3-inch Si/SiGe heterostructure wafer grown by chemical vapor deposition. Electrons are confined to an 8.6~nm thick isotopically purified $^{28}\text{Si}$ layer (800~ppm $^{29}\text{Si}$) with a 41.2~nm $\text{Si}_{0.71}\text{Ge}_{0.29}$ spacer layer capped with approximately 2~nm Si layer. All photolithography is performed using a Nikon I-line stepper (NSR-2005i8A) using positive tone (SPR-955) or image reversal (AZ-5214E) photoresists. Positive tone resist is used in subtractive processes where dielectrics or metals are removed via plasma or wet chemical etching. We find for wet etching an additional post development bake at 110~C for 90~s greatly improves adhesion. Metal lift-off steps are done using image reversal resist due to the $\approx0.5~\mu\text{m}$ undercut after development. Wafers are cleaned after each photolithography step by stripping in 75~C 1165 Remover for 30~minutes followed by 5 minutes with sonication in acetone, IPA,  and DI water. For lift off, we use 75~C 1165 Remover for 30-60~minutes followed by sonication for 5~minutes in two acetone rinses, IPA, and DI water. The sample is then put into a spin rinse dryer followed by a subsequent 5~minute plasma cleaning treatment in a downstream oxygen plasma asher. 

Dopants (P31+) are implanted in $5 \times 10 \mu\text{m}^2$ areas defined by vias in a 300~nm thick $\text{SiO}_2$ hard mask layer grown by plasma enhanced chemical vapor deposition (PEVCD) in a Plasma-Therm 73. To create a more uniform profile of dopants from the surface down to the quantum well, two ion implant energies were used. After ion implant, the wafers are annealed at 700~C for 30~s in a forming gas environment to activate the dopants. The hard mask is stripped in a 20:1 buffered oxide etch (BOE) solution until the surface becomes hydrophobic. The wafer is rinsed and cleaned in a downstream oxygen plasma asher. A second 20:1 BOE dip is done to remove any residual dopants or hard mask oxide that may have landed on the surface during the first strip. Next, two 80~nm tall, $50\times100~\mu\text{m}^2$ area mesa regions on each die are patterned and etched using a $\text{CHF}_3/\text{O}_2$ plasma in a Plasma-Therm 790 reactive ion etcher (RIE). Reflow of the photoresist after patterning creates a $60\deg$ sloped sidewall on the mesa region from pattern transfer of the sloped photoresist during the RIE etch facilitating step coverage for thin films onto the mesa.

Next we form a high-yield, ultrathin (1.7~nm), low noise $\text{SiO}_2$ gate dielectric using boiling $\text{HNO}_3$ acid. We first clean the wafer in a 80~C, 96\% sulfuric acid for ten minutes to remove residual hydrocarbons or metals on the surface of the wafer from the previous processing steps. Next, a strip of the uncontrolled surface oxide is done using a 20:1 BOE solution for 120~s. Thorough rinsing of the wafer is done by rapidly flushing the sample with flowing DI water for 60~s and dried using filtered $\text{N}_2$ with care to ensure the wafer is visually dry and that the surface were hydrophobic during the DI rinse. It is critical no hydroflouric acid remains as HF + $\text{HNO}_3$ form a powerful silicon etchant that can destroy the sample. After drying, the wafer is immediately submerged in a boiling azeotropic (68 wt~\%) $\text{HNO}_3$ for 10 minutes. Next, the wafer is annealed for 30 minutes in a 400~C forming gas environment to passivate trapped charge and dangling bonds with hydrogen. Immediately after, a 17.5~nm $\text{Al}_2\text{O}_3$ is deposited across the wafer using atomic layer deposition which serves as the field oxide over the implant region and protects the gate oxide from subsequent processing. The ALD  is annealed in a SAMCO UV-ozone stripper for 15 minutes to reduce fixed charge in the material. Except for the capping layer on the mesas, the $\text{Al}_2\text{O}_3$ is then removed using 20:1 BOE to minimize any dielectric loss to the superconducting resonator. Next, vias in the $\text{Al}_2\text{O}_3$ layer implant regions are etched using 70~C $\text{HPO}_3$ solution (Transetch-N) from Transene Corporation. After patterning with negative resist, ohmic contact is then formed by using an \textit{in situ} $\text{Ar}^+$ ion mill with subsequent deposition and lift off of a 5~nm Ti, 45~nm Pd metal layer using a home built e-beam metal evaporator. A subsequent pattern and lift off of the dot gate interconnect layer is done in the same evaporator with 3~nm Ti, 12~nm Pd.

The next set of steps create the low impedance buried coplanar waveguides necessary to maintain high Q resonator in the presence of the quantum dot gate stack. We deposit 90~nm niobium on the wafer using DC magnetron sputtering. The base layer metal is patterned and etched using a $\text{BCl}_3/\text{Cl}_2$ plasma in a Plasma-Therm 770 inductively coupled plasma (ICP) etcher with the mesa regions covered by niobium to act as an etch stop for the subsequent dielectric processing. Next, we grow by PECVD a 200~nm thick $\text{SiO}_2$ dielectric layer at 250~C across the whole wafer. The layer is then etched with a sloped sidewall using the $\text{CHF}_3/\text{O}_2$ RIE etch everywhere except over the dot bias leads and parallel plate capacitor for the center pin DC tap. The niobium surfaces are further cleaned by a 20:1 BOE etch to remove any residual $\text{SiO}_2$ not removed by the RIE etching. Next, a 90~nm counter electrode layer of niobium is sputtered and lifted off using an \emph{in situ} ion mill prior to deposition to remove oxide and facilitate ohmic contact between metal layers. Landing pads for the bump bonding are formed using in situ ion mill and deposition of 10~nm Ti, 25~nm Pd, transfer to a different metal evaporator (Lesker PVD-75), and deposition of 20~nm Ti, 50~nm Pt, 100~nm Au. Darkening of the niobium counterelectrode layer was observed after lift off, likely due to chemical reaction between the niobium and photoresist from the high chamber temperature during the Pt deposition.

The protective Nb capping layer over the mesa regions is then removed using an SF6 etch which is highly selective of Nb over Pd and $\text{Al}_2\text{O}_3$. The $\text{Al}_2\text{O}_3$ is then removed off the $\text{SiO}_2$ with high selectivity by a 70~C $\text{HPO}_3$ etch for 180~s everywhere on the mesa except for the implant regions and under the interconnects. Profilometry is used to confirm etch through using a symmetric grid test points across the wafer. The wafer is then diced into pieces with a protective photoresist layer spun on prior to dicing. Electron beam lithography of the nano-gates is done on the pieces using an Elionix ELS-100 with PMMA 495 A4 as the lift off resist mask. Typical layer doses are 1100-1350 $\mu\text{C}/\text{cm}^2$ for both single and multilayer structures. We note in multilayer structures, higher doses compared to dose test structures are required for layers 2 and 3, likely due to differences in the backscattered electron energies or densities in the presence of the lower aluminum layers. E-beam evaporation of aluminum is done at rates varying between 0.3 - 1.0 $\AA/s$ depending on the fabrication run. Faster depositions produce higher quality aluminum films in terms of purity due to reduced water contamination during evaporation, but intermittently have larger grain size than slower grown films. For multi-layer aluminum gate stack devices, the aluminum layers are 25~nm, 40~nm, and 65~nm thick with 10 minutes of oxygen plasma ashing done after each lift off step to clean and further oxidize the aluminum for high yield gate to gate isolation. For the single layer device discussed in the main text we used the first layer deposition parameters of the multilayer devices. 

Fabrication of the superconducting resonators began with physical vapor deposition of 50 nm TiN films on high-resistivity ($>10~\text{k}\Omega$) 8" silicon wafers that were treated with an SC1/SC2 megasonic clean. The resonators were defined using 248 nm lithography and a chlorine-based plasma etch, which resulted in approximately 220 nm trenching into the silicon surface. An underbump metalization layer, comprising 20 nm of titanium, 50 nm of platinum, and 100 nm of gold, was evaporated and patterned using a liftoff step. Finally, an indium film approximately 10 um thick was thermally evaporated and patterned using liftoff into pillars with a diameter of 15 um.

The silicon qubit chip and the superconducting resonator chip were bonded together using a commercial thermo-electric compression bonding system with a post-bond lateral accuracy of approximately 1 um. Bonding was performed at 105 degrees C with 3000 g of force, resulting in a post-bond spacing between chips of approximately 3 um.

\subsection{Experimental Wiring Diagram}

\begin{figure*}[ht]
\includegraphics[width=2\columnwidth]{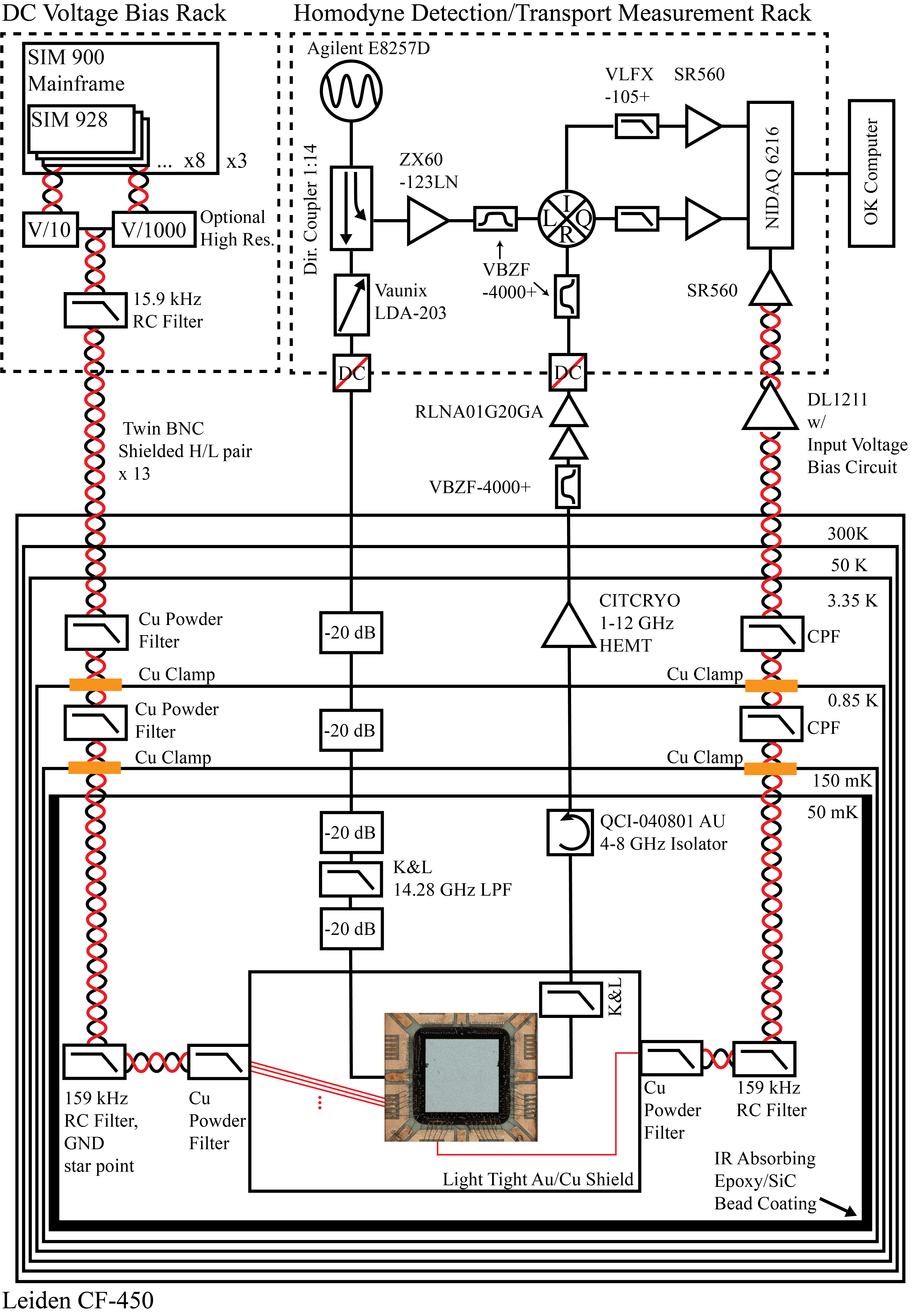}
\caption{\label{fig:wiring} Schematic of the fridge wiring.}
\end{figure*}

\subsection{Additional Charge Noise Data}
\begin{figure*}[ht]
\includegraphics[width=2\columnwidth]{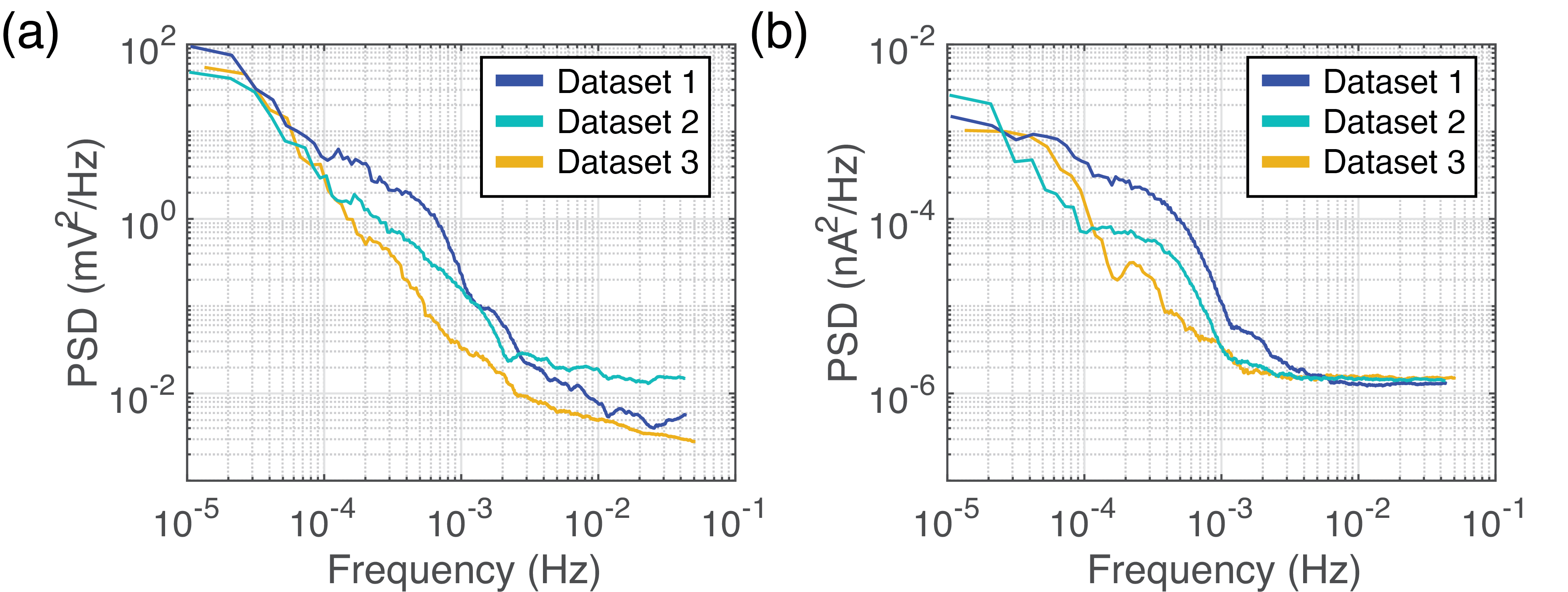}
\caption{\label{fig:AddCN} (a) Comparison of the extracted offset voltage noise using the peak monitoring method over the course of three days for the quantum dot formed under B2. These data show similar amplitudes and power law trends over the measured range. (b) The corresponding peak current power spectrum from the same datasets. We observe Lorentzian features in the noise spectrum of the peak current below 1~mHz.}
\end{figure*}

Figure \ref{fig:AddCN} provides additional (smoothed) very low frequency noise data from the dot formed under B2. We observe the same general trend over the course of three days. Offset current noise is also extracted and shows some data sets exhibit large Lorentzian features in conjunction with the general $1/f^2$ power law. We note higher peak location noise coincides with the apparent amplitude of these features.

\subsection{Additional SNR Data}
\begin{figure*}[ht]
\includegraphics[width=2\columnwidth]{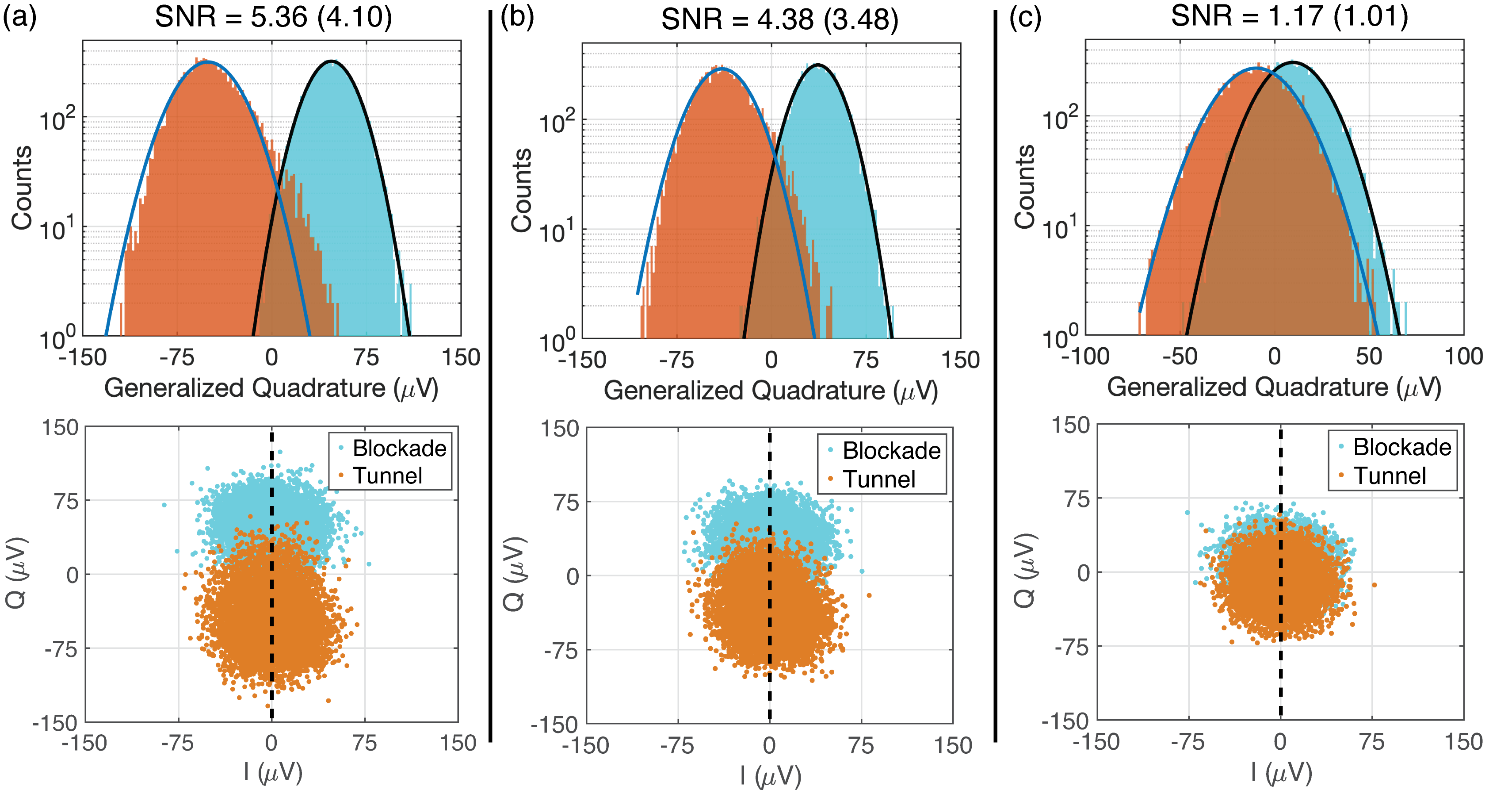}
\caption{\label{fig:SNR} (a) Signal to noise ratio data with an on-chip probe power of $P_{\text{in}} = -95~\text{dBm}$. (b) Signal to noise ratio data with an on-chip probe power of $P_{\text{in}} = -98~\text{dBm}$. (c) Signal to noise ratio data with an on-chip probe power of $P_{\text{in}} = -105~\text{dBm}$. For all datasets we observe a larger standard deviation in the tunneling peak compared to the blockade peak. For completeness, the SNR defined using the larger variation is given in parentheses for each dataset.}
\end{figure*}

Figure \ref{fig:SNR} (a) - (c) provides additional dispersive readout SNR data as a function of the on-chip probe power $P_{\text{in}}$.


\bibliography{aipsamp}

\end{document}